%% file: main.tex
\PassOptionsToPackage{table}{xcolor}
\documentclass[sigconf]{acmart}

\AtBeginDocument{%
  }

\setcopyright{none}
\copyrightyear{2026}
\acmYear{2026}
\acmConference[JCDL '26]{Proceedings of the ACM/IEEE Joint Conference on Digital Libraries}{October 13--16, 2026}{Dallas, Texas, USA}
\acmBooktitle{Proceedings of the ACM/IEEE Joint Conference on Digital Libraries (JCDL '26), October 13--16, 2026, Dallas, Texas, USA}
\acmDOI{}
\acmISBN{}

\usepackage{booktabs}
\usepackage{array}
\usepackage{multirow}
\usepackage{tabularx}
\usepackage{enumitem}
\usepackage{placeins}
\usepackage{xspace}
\definecolor{TableBlue}{HTML}{EAF3FA}
\definecolor{TableBlueDark}{HTML}{D8EAF8}
\definecolor{TableGreen}{HTML}{EAF6EA}
\definecolor{TableGreenDark}{HTML}{DDEFD8}
\definecolor{TableRed}{HTML}{F8E6E6}
\definecolor{TableOrange}{HTML}{FFF1D6}
\definecolor{TableGray}{HTML}{F2F2F2}
\definecolor{TableRule}{HTML}{6E6E6E}
\newcolumntype{Y}{>{\raggedright\arraybackslash}X}
\newcommand{\tableheader}{}
\newcommand{\tablesubhead}[2]{\multicolumn{#1}{@{}l}{\textit{#2}}}

\newcommand{\curatability}{curatability\xspace}
\newcommand{\hf}{Hugging Face\xspace}
\newcommand{\llm}{LLM\xspace}
\newcommand{\llms}{LLMs\xspace}

\begin{document}

\title{Visible but Not Yet Curatable: Characterizing the Curatability of Compact and Derived Open LLM Artifacts}

\author{Yiyi Lu}
\authornote{Yiyi Lu and Yilai Qian contributed equally to this work.}
\email{yl996@duke.edu}
\affiliation{%
  \institution{Duke Kunshan University}
  \city{Kunshan}
  \state{Jiangsu}
  \country{China}}

\author{Yilai Qian}
\authornotemark[1]
\email{yq120@duke.edu}
\affiliation{%
  \institution{Duke Kunshan University}
  \city{Kunshan}
  \state{Jiangsu}
  \country{China}}

\author{Yucheng Jin}
\authornote{Corresponding author.}
\email{yj232@duke.edu}
\affiliation{%
  \institution{Duke Kunshan University}
  \city{Kunshan}
  \state{Jiangsu}
  \country{China}}

\renewcommand{\shortauthors}{Lu et al.}

\begin{abstract}
Open Large Language Model (LLM) research increasingly produces compact and derived artifacts, such as adapters, quantized checkpoints, merged models, and distilled variants, that are distributed across papers, model hubs, model cards, code repositories, and release statements. Although these artifacts are publicly visible, digital libraries often lack sufficient evidence to identify, preserve, and cite them as coherent scholarly objects. We introduce a framework that conceptualizes curatability as a record-level property of distributed scholarly records and operationalizes it through four evidence dimensions: artifact identity, scholarly linkage, upstream evidence, and release assets.
Guided by this framework, we conduct the updated collection-scale characterization of open \llm{} curatability using a May~2026 snapshot of 191,375 public Hugging Face repositories and a core corpus of 2,214 scholarly papers. Our results reveal a pronounced visibility-to-curatability funnel. While 90.7\% of paper records contain at least one useful curation signal, only 18.1\% combine usable upstream evidence with concrete release evidence, and only 6.1\% provide sufficiently coordinated evidence to support high-curatability records.
Based on these findings, we derive a minimal seven-field curatable record and complementary responsibilities for model hubs, scholarly indexes, and digital libraries, providing practical guidance for improving the preservation and bibliographic control of open \llm{} artifacts. The code and results are available at:
\url{https://github.com/AndyLu666/VNYC-Open_Source_LLM_Study}.
\end{abstract}



\maketitle

\input{sections/01_introduction}
\input{sections/02_problem}
\input{sections/03_related_work}
\input{sections/04_data_methods}
\input{sections/05_results}
\input{sections/06_discussion}
\input{sections/07_limitations_conclusion}

\bibliographystyle{ACM-Reference-Format}
\bibliography{references}

\end{document}

%% file: sections/01_introduction.tex
\section{Introduction}

Digital libraries increasingly curate scholarly outputs that extend beyond the traditional narrative article to encompass diverse research objects, including datasets, software, scripts, trained models, evaluation resources, and external repositories \cite{wilkinson2016fair,smith2016softwarecitation,barker2022fair4rs,soilandreyes2022rocrate}. Preserving these heterogeneous resources is essential for ensuring the transparency, reproducibility, and long-term accessibility of computational research. While digital library infrastructures have progressively adapted to software and data, the rapid growth of open Large Language Model (\llm) research introduces a new class of computational artifacts whose curation extends beyond the scope of existing bibliographic practices.

Open \llm research follows a collaborative development paradigm in which model weights, training code, datasets, and evaluation resources are publicly released and continuously extended by the community. The resulting scholarly outputs, which we collectively refer to as \textit{open \llm artifacts}, include not only foundation models but also numerous derived computational objects such as LoRA adapters, quantized checkpoints, merged models, distilled variants, and deployment packages. Unlike conventional research outputs, these artifacts are inherently relational: their functionality, provenance, and scientific meaning depend on explicit connections to upstream models, training resources, release assets, and associated publications.

From a digital library perspective, open \llm artifacts therefore constitute distributed research objects rather than self-contained publications or repositories. A single paper may introduce multiple interdependent artifacts whose evidence is dispersed across article text, Hugging Face repositories, model cards, source code repositories, platform metadata, version histories, and release notes. Although these records are publicly accessible, their metadata rarely specifies which artifact is the intended scholarly contribution, whether a repository represents an official release or a derivative checkpoint, or how individual artifacts relate to upstream dependencies. Consequently, artifact identity, provenance, scholarly linkage, and release evidence remain fragmented across independently maintained platforms. The preservation challenge is therefore not merely locating artifacts, but assembling sufficient contextual evidence to identify, cite, preserve, and safely reuse them as coherent scholarly objects.

Existing research provides important foundations for documenting computational artifacts. Work on FAIR principles, software citation, research-object packaging, model cards, datasheets, and model openness frameworks consistently argues that computational research requires structured metadata beyond simple resource availability \cite{gebru2021datasheets,mitchell2019modelcards,pineau2021reproducibility,bommasani2023fmi,white2024mof,salsabil2025context}. Complementary studies of model hubs and open-source AI communities further demonstrate how models evolve through continual reuse and adaptation \cite{jiang2023ptmreuse,taraghi2024modelreuse,osborne2024aicommunity}. However, most prior work examines only a single component of the scholarly record, such as papers, repositories, model cards, datasets, or platform-specific metadata, in isolation. As a result, little empirical evidence exists regarding how these heterogeneous records collectively support the identification, preservation, and bibliographic control of open \llm artifacts at collection scale.

To address this gap, we introduce \curatability, a record-level measure of whether publicly available evidence is sufficient for digital library curation. A curatable record should provide enough information to identify the target artifact, establish its scholarly and technical relationships, and expose the release assets required for inspection and preservation. Specifically, we characterize curatability using four complementary evidence dimensions: artifact identity, record-supported scholarly linkage, upstream model evidence, and release assets.

Using a fixed May~2026 snapshot, we analyze 191,375 public \hf{} model repositories together with 2,729 scholarly paper records, yielding a core analysis set of 2,214 papers. Our measurement pipeline links repositories and publications by integrating normalized repository metadata, paper-side links, provenance signals, release-package information, and cross-platform alignment evidence. The resulting measurements are further validated through expert audits and sensitivity analyses.

Our cross-system analysis reveals a pronounced \textit{visibility-to-curatability funnel}. While 90.7\% of paper records expose at least one useful metadata signal, far fewer provide the coordinated evidence required for reliable curation. Only 5.6\% of papers contain direct links to Hugging Face repositories. Usable provenance combined with executable release assets appears in 18.1\% of records, and only 6.1\% simultaneously provide explicit release-package evidence together with direct paper--repository alignment. These findings suggest that the principal preservation challenge is not the absence of publicly released artifacts, but the fragmentation of the contextual evidence required to identify, attribute, and preserve them across distributed scholarly records.

This study makes three contributions:

\begin{itemize}[leftmargin=*]

    \item \textbf{A digital-library framework for curatability.} We introduce \curatability as a record-level concept for assessing whether distributed public records provide sufficient evidence for preserving, identifying, and citing open \llm artifacts. We operationalize curatability through four complementary evidence dimensions: artifact identity, scholarly linkage, upstream evidence, and release assets.

    \item \textbf{The first collection-scale characterization of open \llm curatability.}
    Using 191,375 Hugging Face repositories and 2,214 scholarly papers, we quantify how artifact identity, provenance, scholarly linkage, and release evidence co-occur across public records, revealing a pronounced visibility-to-curatability funnel.

    \item \textbf{Actionable guidance for digital library curation.}
    We derive a minimal seven-field curatable record together with complementary responsibilities for model hubs, scholarly indexes, and digital libraries, providing practical recommendations for improving preservation and bibliographic control of open \llm artifacts.
\end{itemize}

\begin{figure}[t]
\centering
\includegraphics[width=\linewidth]{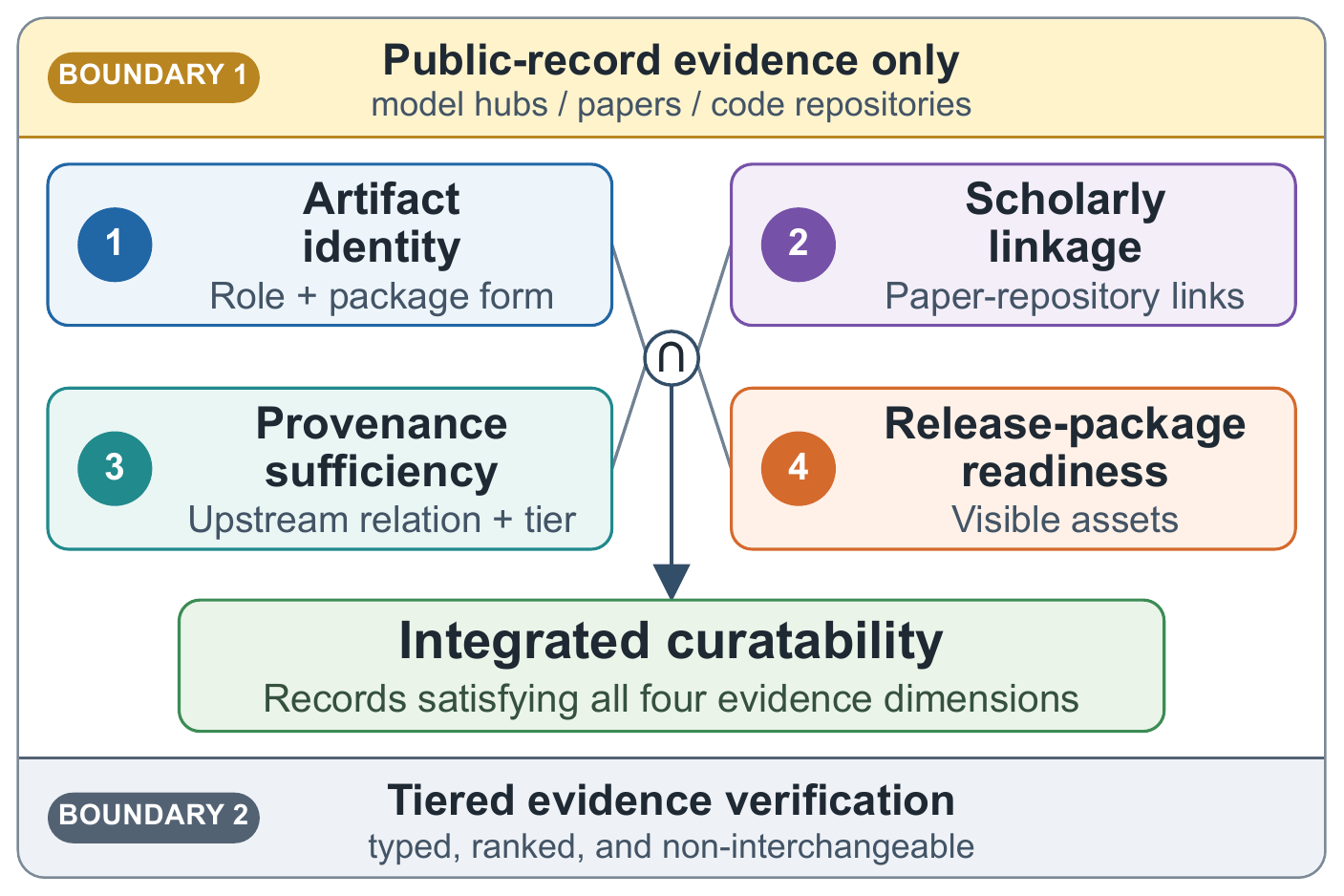}
\caption{\textbf{Curatability framework for open \llm artifacts.} Curatability is operationalized through four evidence dimensions—artifact identity, scholarly linkage, provenance evidence, and release assets—under two operational constraints: reliance on public metadata and tiered evidence interpretation.}
\Description{A framework diagram showing four evidence dimensions for open LLM artifact curatability: artifact identity, scholarly linkage, provenance evidence, and release assets. The dimensions are evaluated using public metadata and tiered evidence interpretation.}
\label{fig:framework}
\end{figure}

%% file: sections/02_problem.tex
\section{Curatability Framework}
\label{sec:framework}

To operationalize the curation of open \llm artifacts, we build upon established digital library principles including FAIR research objects, software citation, provenance models, and model documentation \cite{wilkinson2016fair,smith2016softwarecitation,soilandreyes2022rocrate,mitchell2019modelcards,gebru2021datasheets,moreau2013prov,pineau2021reproducibility,salsabil2025context}. Rather than evaluating individual repositories or papers in isolation, our framework assesses whether distributed public records collectively provide sufficient evidence for digital library curation. We refer to this property as \textit{curatability}.

\begin{figure*}[!tbp]
    \centering
    \includegraphics[width=\textwidth]{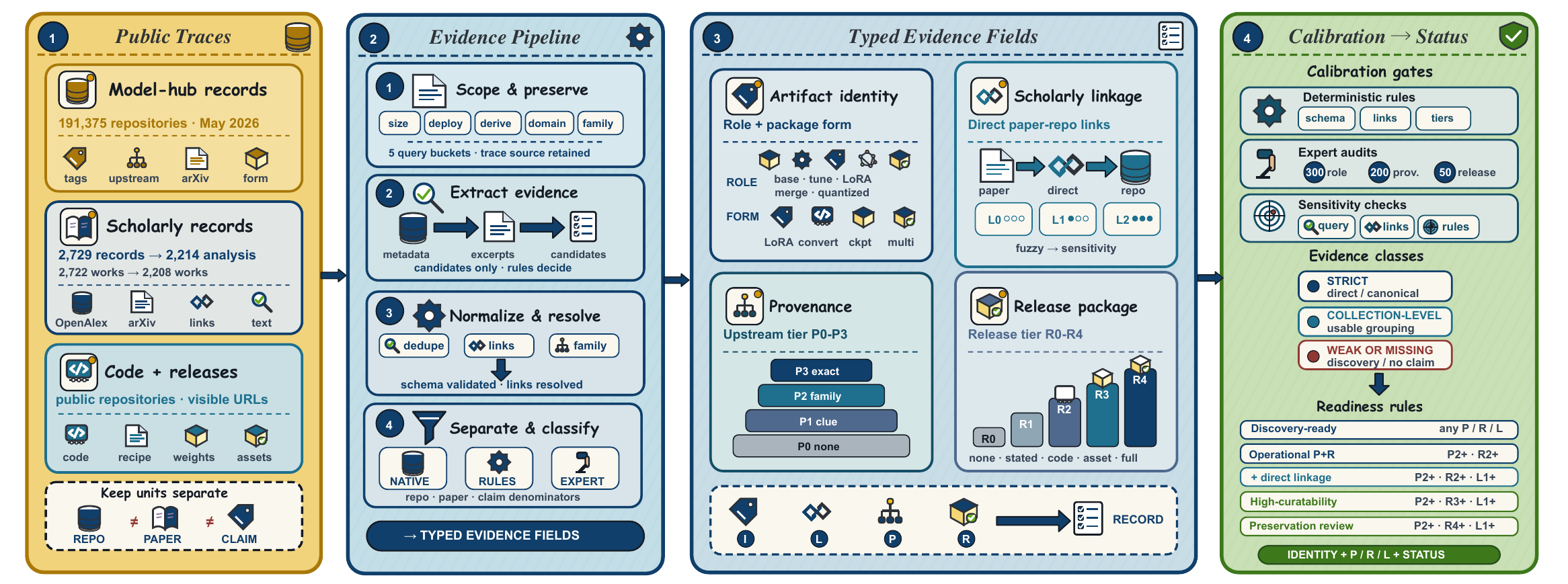}
  \caption{\textbf{Systemic Evidence Integration and Curation Calibration Workflow.} Evaluating a corpus of 191,375 Hugging Face repositories and 2,214 paper records, the pipeline processes scattered traces through four sequential stages: \textbf{Panel 1} ingests multi-platform public traces; \textbf{Panel 2} extracts and normalizes the evidence; \textbf{Panel 3} structures these signals into four typed fields; and \textbf{Panel 4} calibrates evidence strength to output final curation-readiness decisions.}
\label{fig:pipeline}
\Description{A workflow diagram mapping the transition from public traces to curation decisions across four panels evaluating 191,375 repositories and 2,214 paper records.}
\end{figure*}

As shown in Figure~\ref{fig:framework}, curatability is operationalized through four complementary evidence dimensions:

\begin{itemize}[leftmargin=*]
    \item \textbf{Artifact identity} determines whether the target scholarly artifact can be uniquely identified (e.g., foundation model, adapter, merged model, quantized checkpoint, or deployment package).

    \item \textbf{Scholarly linkage} captures explicit connections among papers, repositories, creators, licenses, and task contexts.

    \item \textbf{Provenance evidence} evaluates whether relationships to upstream models are explicitly documented through traceable evidence rather than inferred from contextual mentions \cite{moreau2013prov,davidson2008provenance,herschel2017survey}.

    \item \textbf{Release assets} assess whether executable research outputs, such as model weights, source code, training scripts, or evaluation resources, are publicly available to support inspection and preservation \cite{dodge2019show,pineau2021reproducibility,akella2023effort}.
\end{itemize}

A record is considered curatable only when these evidence dimensions collectively support reliable identification, attribution, and preservation. To ensure consistent assessment, we impose two operational constraints: (1) only publicly observable metadata are considered, and (2) evidence is interpreted according to a tiered hierarchy, distinguishing repository metadata, paper--repository links, provenance statements, model cards, and release assets rather than treating them as equivalent evidence sources \cite{mitchell2019modelcards,pepe2024hfdocs,kadasi2025modelhubs,salsabil2025context}.

These evidence dimensions provide the operational definitions used throughout the remainder of the paper, guiding metadata extraction (Section~4), quantitative analysis (Section~5), and the design of the minimal curatable record (Section~6).

%% file: sections/03_related_work.tex
\section{Related Work}
\label{sec:related}

\subsection{Computational Research Objects}

Digital library and open-science frameworks increasingly recognize datasets, software, code repositories, URLs, and packaged research objects as first-class components of the scholarly record. Rather than treating computational artifacts as simple downloadable resources, these frameworks emphasize persistent identifiers, authorship, licensing, dependencies, provenance, and contextual metadata to support long-term preservation and reuse~\cite{wilkinson2016fair,datacitation2014joint,smith2016softwarecitation,lamprecht2020fairsoftware,katz2021freshfairsoftware,barker2022fair4rs,soilandreyes2022rocrate}. Studies of scholarly URLs similarly demonstrate that preserving hyperlinks alone is insufficient because surrounding textual context is often required to interpret their scholarly role and support curation decisions~\cite{salsabil2025context}.

These studies establish the importance of rich metadata for computational research objects but largely assume that the object being curated is already identifiable. Open \llm artifacts challenge this assumption because a single scholarly contribution may consist of multiple derived artifacts distributed across papers, model hubs, repositories, and documentation. This motivates evaluating not only resource availability but also whether distributed records collectively provide sufficient evidence for digital library curation.

\subsection{Documentation and Release of Open \llm Artifacts}

Research on computational reproducibility emphasizes that public availability alone does not guarantee meaningful reuse. Reproducibility depends on the coordinated release of source code, executable environments, model weights, training scripts, evaluation assets, and supporting documentation~\cite{stodden2014implementing,dodge2019show,akella2023effort,pineau2021reproducibility}. Building on this perspective, model cards, datasheets, data cards, Croissant, and related documentation frameworks promote structured descriptions of machine-learning assets, while openness frameworks advocate reporting multiple components of model development rather than weights alone~\cite{mitchell2019modelcards,gebru2021datasheets,pushkarna2022datacards,crisan2022interactive,shen2022modelcardtoolkit,akhtar2024croissant,bommasani2023fmi,white2024mof}.

Empirical analyses further show substantial variation in documentation practices across Hugging Face repositories and open-source AI communities, particularly in reporting licensing, datasets, model properties, and reuse guidance~\cite{liang2024modelcards,pepe2024hfdocs,kadasi2025modelhubs,jiang2023ptmreuse,taraghi2024modelreuse,osborne2024aicommunity}. Recent work has also assembled model-card metadata for multidisciplinary analysis and audited license integrity across Hugging Face--GitHub supply chains~\cite{suryani2025modelcardmetadata,jewitt2026permissivewashing}. These studies measure repository metadata, documentation, reuse, and license integrity. Our analysis instead asks whether paper, repository, upstream-relation, and release-asset evidence co-occur as a curatable cross-system scholarly record.

\input{tables/table_corpus_construction}

\subsection{Provenance and Cross-platform Scholarly Records}

Provenance models provide formal mechanisms for describing how scholarly objects are created, transformed, and derived~\cite{moreau2013prov,davidson2008provenance,herschel2017survey}. The W3C PROV model represents lineage through entities, activities, and agents, while digital-library applications demonstrate how provenance supports transparency and preservation in document workflows~\cite{moreau2013prov,rauch2025ontextract}. These principles are particularly relevant to open \llm artifacts, whose scientific meaning often depends on explicit relationships to upstream foundation models, fine-tuning checkpoints, adapters, distillation teachers, or evaluation baselines~\cite{jiang2023ptmreuse,taraghi2024modelreuse,white2024mof}.

In practice, however, these relationships are frequently distributed across publications, model hubs, repository metadata, and documentation, with many records relying on family names or contextual references rather than explicit machine-readable provenance. FAIR Signposting demonstrates how typed Web links can expose identifiers, metadata, content, authorship, and licensing for distributed digital objects~\cite{vandesompel2024fairsignposting}. We extend this typed-relation perspective by measuring whether scholarly linkage co-occurs with usable upstream and release evidence for open \llm artifacts. Our unit of analysis is therefore the linked cross-platform scholarly record rather than an individual paper or repository.

%% file: tables/table_corpus_construction.tex
\begin{table*}[!tbp]
\centering
\caption{Corpus construction summary. Counts are record counts used for measurement; detailed query strings, source files, and derived tables are provided in the public artifact package.}
\label{tab:corpus-construction}
\small
\setlength{\tabcolsep}{5pt}
\renewcommand{\arraystretch}{1.13}
\begin{tabularx}{0.98\textwidth}{@{}p{0.18\textwidth}p{0.24\textwidth}Y >{\raggedleft\arraybackslash}p{0.12\textwidth}@{}}
\toprule
\textbf{Layer} & \textbf{Input source} & \textbf{Scope / filter} & \textbf{Records} \\
\midrule
\rowcolor{TableBlue}
\tablesubhead{4}{Repository snapshot} \\
\cmidrule(lr){1-4}
Model-hub records & Public \hf model repositories & Targeted compact/derived query buckets; duplicate repository ids consolidated & 191,375 \\
\midrule
\rowcolor{TableBlue}
\tablesubhead{4}{Scholarly paper records} \\
\cmidrule(lr){1-4}
Paper records & OpenAlex, arXiv, venue lists & Candidate compact/derived \llm studies after source merge & 2,729 \\
Unique works & Deduplicated paper metadata & DOI/arXiv/title-level consolidation & 2,722 \\
\rowcolor{TableGreen}
\textbf{Main analysis set} & \textbf{Filtered paper records} & \textbf{Artifact-centered studies; app-only, benchmark-only, and generic-utility papers excluded} & \textbf{2,214} \\
Main unique works & Deduplicated main set & Work-level counterpart of main analysis records & 2,208 \\
\bottomrule
\end{tabularx}
\renewcommand{\arraystretch}{1.0}
\end{table*}

%% file: sections/04_data_methods.tex
\section{Data and Measurement Design}
\label{sec:methods}
Figure~\ref{fig:pipeline} organizes the measurement design into four components that also structure this section. We first assemble public traces from model hubs, scholarly records, and code repositories. We then extract and normalize evidence while preserving its source and unit of analysis. The resulting signals are represented as artifact identity, scholarly linkage, provenance sufficiency, and release-package readiness. Expert audits calibrate judgment-sensitive boundaries, and sensitivity checks test the deterministic rules used to assign curatability status. Repository, paper, and artifact-claim denominators remain separate throughout. The public artifact package provides the queries, schemas, intermediate tables, and scripts.

\subsection{Public Traces}

\paragraph{Model-hub records.}
We collected a targeted May 2026 snapshot of 191,375 public \hf model repositories. The five query buckets covered size or compactness, deployment format, derivation or tuning, domain-related characteristics including language and task, and model family names. Duplicate repository identifiers were consolidated. We retained core repository metadata, comprising tags, upstream fields, scholarly references, task and family labels, license signals, and domain descriptors. Because the collection is query based, the estimates describe this scoped snapshot rather than all \hf repositories. Query-bucket sensitivity tests assess scope variation.
\input{tables/table_audit_wilson_ci}
\input{tables/table_score_ladder}
\paragraph{Scholarly records.}
We merged OpenAlex, arXiv, and venue-specific lists, yielding 2,729 paper records and 2,722 unique works after deduplication. Filtering produced a main set of 2,214 records and 2,208 works. Eligible papers introduced, adapted, compressed, quantized, distilled, fine-tuned, evaluated, or deployed a compact or derived model as the primary research object. We excluded application-only pipelines, benchmark-only studies, and papers that used \llms only as generic utilities. The scope is operational rather than based on parameter count alone. This scope encompasses both small-footprint and deployment-oriented models, alongside derived artifacts whose interpretation depends on upstream models or released assets. 

\paragraph{Code and release traces.}
We collected public URLs and asset statements from papers, model cards, and matched repositories. Each trace retains its source and local context.

\paragraph{Units of analysis.}
Repository-level percentages use the 191,375-record snapshot, and paper-level percentages use the 2,214 analysis-set records. Artifact-level language refers to claims supported by these records rather than a census of unique artifacts. One paper may describe several artifacts, and one repository may contain several package forms.

\subsection{Evidence Pipeline}

\paragraph{Scope and preserve.}
Each trace is stored with its source platform, record identifier, query bucket, and evidence location. This keeps repository metadata, paper statements, model-card fields, and code records distinct.

\paragraph{Extract evidence.}
Structured metadata and URLs are parsed directly. For unstructured paper text, an LLM extracts candidate fields and short supporting excerpts. These outputs are candidate evidence only, and final labels are assigned after rule-based validation.

\paragraph{Normalize and resolve.}
We deduplicate works and repositories, normalize URLs and model names, match curated family inventories, and resolve paper-repository links. Deterministic parsers take precedence when structured evidence is available.

\paragraph{Separate and classify.}
Native fields are observed public metadata. Rule-derived fields are generated by documented scripts. Judgment-sensitive cases are passed to the audits in Section~4.4. Parser rules, prompt templates, and intermediate mappings are included in the artifact package.

\subsection{Typed Evidence Fields}

\paragraph{Artifact identity.}
Repository metadata and tags support two fields. Artifact role distinguishes base checkpoints, fine-tunes, adapters or LoRA, merges or mixed records, and quantized or deployment packages. Package form distinguishes adapter-only, quantized, converted, checkpoint-complete, code-only, and multi-asset distributions. Artifact Role identifies the object, while package form identifies how it is distributed.

\paragraph{Scholarly linkage.}
Paper-side \hf links, code-hosting links, repository-side arXiv references, exact model mentions, and model-card backlinks remain separate signals. $L=0$ indicates no direct relation signal, $L=1$ one direct alignment signal, and $L=2$ corroboration by multiple direct signals. Unlike $P$ and $R$, $L$ functions primarily as a \emph{relation gate}: $L\geq1$ records whether a repository or released asset can be connected bibliographically to the focal paper, while $L=2$ records corroboration rather than a stronger curation construct. Fuzzy family or size matches are used only in sensitivity analysis and do not satisfy $L\geq1$.

\paragraph{Provenance sufficiency.}
Each paper record receives its strongest supported tier. $P=0$ indicates no usable upstream signal, $P=1$ a contextual family clue, $P=2$ inventory-backed family evidence, and $P=3$ explicit or canonical upstream evidence. A link contributes only when the paper or repository exposes an upstream relation. $P=2$ supports collection-level grouping, while $P=3$ is required for an explicit lineage claim.

\paragraph{Release-package readiness.}
$R=0$ indicates no release signal, $R=1$ a stated or incomplete release, $R=2$ code, scripts, or a recipe, $R=3$ a checkpoint, adapter, data, or evaluation asset, and $R=4$ a multi-asset package or full recipe. The scores measure visible evidence for inspection and preservation triage rather than executable reproducibility or role-specific package sufficiency.

\subsection{Calibration and Curatability Status}

The final component of Figure~\ref{fig:pipeline} maps typed fields to status through rule-based thresholds. Expert audits calibrate judgment-sensitive coding boundaries. A balanced role audit reviewed 300 repositories, with 60 per predicted role. An upstream-evidence audit reviewed 200 paper records and separated strict unique-primary support from usable family-level support. A release audit reviewed 50 URL or asset rows. One study-team expert reviewed each sampled row against the retained public evidence and recorded a decision, supporting excerpt, and audit note under the shared coding rubric; boundary decisions were then incorporated into the deterministic rules. This single-expert adjudication protocol evaluates support for rule-derived labels, and its outcomes are summarized as support rates with Wilson 95\% intervals. The artifact package provides the coding guide, score rubric, threshold rationale, and curated boundary cases. These audits calibrate coding boundaries rather than estimate corpus prevalence. We also tested query-bucket scope, linkage-signal composition, and alternative thresholds.

\input{tables/table_key_results}

The rubric is based on the curation functions defined in Section~2 rather than on the observed score distribution. Artifact identity is assessed separately through role and package-form evidence because these categories are nominal rather than ordinal. The paper-level readiness thresholds therefore combine provenance, release, and linkage. $P$ and $R$ are ordinal evidence ladders, whereas $L$ is a relation gate with a corroborated state; their values are not comparable across fields. $P\geq2$ is the minimum for collection-level upstream grouping, $R\geq2$ for inspecting a concrete technical release, and $L\geq1$ for direct cross-record alignment. $R\geq3$ requires an asset beyond code or a recipe, and $R\geq4$ identifies a multi-asset package or full recipe. Explicit lineage claims still require $P=3$.

A \textit{discovery-ready} record contains at least one nonzero provenance, release, or linkage signal. The \textit{operational P+R} threshold requires $P\geq2$ and $R\geq2$. The \textit{operational+link} sensitivity check adds $L\geq1$. The \textit{high-curatability} threshold requires $P\geq2$, $R\geq3$, and $L\geq1$. \textit{Preservation-review} candidates require $P\geq2$, $R\geq4$, and $L\geq1$. These labels describe public-record evidence. They do not certify complete genealogy, role-specific release sufficiency, executable reproducibility, or preservation success.

%% file: tables/table_audit_wilson_ci.tex
\begin{table*}[!tbp]
\centering
\caption{Expert-audit outcomes with Wilson 95\% intervals. Audits calibrate boundary decisions rather than estimate population prevalence.}
\label{tab:audit-wilson-ci}
\small
\setlength{\tabcolsep}{5pt}
\renewcommand{\arraystretch}{1.14}
\begin{tabularx}{0.94\textwidth}{@{}p{0.16\textwidth}p{0.21\textwidth}cp{0.18\textwidth}Y@{}}
\toprule
\textbf{Audit layer} & \textbf{Boundary checked} & \textbf{Count} & \textbf{Estimate (95\% CI)} & \textbf{Audit implication} \\
\midrule
Repository role & Strict role support & 256/300 & \cellcolor{TableGreen}\textbf{85.3\%} [80.9, 88.9] & Artifact-role labels are stable \\
\midrule
Upstream evidence & Strict unique-primary & 56/200 & \cellcolor{TableRed}28.0\% [22.2, 34.6] & Strict lineage claims stay conservative \\
 & Usable family-level & 158/200 & \cellcolor{TableGreen}\textbf{79.0\%} [72.8, 84.1] & Family-level grouping is usable \\
\midrule
Release evidence & URL/asset category retained & 50/50 & \cellcolor{TableGreen}\textbf{100.0\%} [92.9, 100.0] & Release categories are stable \\
\bottomrule
\end{tabularx}
\renewcommand{\arraystretch}{1.0}
\end{table*}

%% file: tables/table_score_ladder.tex
\begin{table*}[!tbp]
\centering
\caption{Operational coding for record-level evidence. $P$ and $R$ are evidence ladders; $L$ is a relation gate, with $L=2$ indicating corroboration. The codes define what a public paper record supports rather than model quality or execution reproducibility.}
\label{tab:score-ladder}
\small
\setlength{\tabcolsep}{5pt}
\renewcommand{\arraystretch}{1.14}
\begin{tabularx}{0.98\textwidth}{@{}cp{0.25\textwidth}p{0.26\textwidth}p{0.20\textwidth}Y@{}}
\toprule
\textbf{Code} & \textbf{Provenance $P$} & \textbf{Release $R$} & \textbf{Relation gate $L$} & \textbf{Curation use} \\
\midrule
0 & No usable upstream signal & No release signal & No direct signal & No supported claim \\
\rowcolor{TableGray}
1 & Weak or contextual family clue & Stated or incomplete release & One direct alignment signal & Discovery cue \\
2 & Inventory-backed family evidence & Code, scripts, or recipe & Multiple direct signals & Grouping / inspection \\
\rowcolor{TableGray}
3 & Explicit or canonical upstream evidence & Checkpoint, adapter, data, or evaluation asset & \textit{not used} & Upstream or asset-level claim \\
4 & \textit{not used} & Multi-asset package or full recipe & \textit{not used} & Preservation-review candidate \\
\bottomrule
\end{tabularx}
\renewcommand{\arraystretch}{1.0}
\end{table*}

%% file: tables/table_key_results.tex
\begin{table*}[!tbp]
\centering
\caption{Headline evidence components and curation thresholds. Counts use 2,214 analysis-set paper records. $P$, $R$, and $L$ denote provenance, release-package, and direct-linkage scores.}
\label{tab:key-results}
\normalsize
\setlength{\tabcolsep}{5.5pt}
\renewcommand{\arraystretch}{1.13}
\begin{tabularx}{0.98\textwidth}{@{}p{0.19\textwidth}Yrrp{0.24\textwidth}@{}}
\toprule
\tableheader
\textbf{Group} & \textbf{Evidence or threshold} & \textbf{Count} & \textbf{Share} & \textbf{Takeaway} \\
\midrule
\rowcolor{TableBlue}
\tablesubhead{5}{Evidence components} \\
\cmidrule(lr){1-5}
Linkage & Matched paper--HF repo & 124 / 2,214 & 5.6\% & Direct relation rare \\
Provenance & Inventory-backed family & 1,009 / 2,214 & 45.6\% & Grouping, not lineage \\
Release & Code/asset evidence ($R\geq2$) & 793 / 2,214 & 35.8\% & Concrete release uneven \\
\midrule
\rowcolor{TableBlue}
\tablesubhead{5}{Curation-readiness thresholds} \\
\cmidrule(lr){1-5}
Discovery-ready & Any useful signal & 2,008 / 2,214 & \cellcolor{TableBlueDark}\textbf{90.7\%} & Broad visibility \\
Operational P+R & $P\geq2, R\geq2$ & 401 / 2,214 & \cellcolor{TableGreenDark}\textbf{18.1\%} & Usable P + release \\
Operational+link & $P\geq2, R\geq2, L\geq1$ & 400 / 2,214 & \cellcolor{TableGreenDark}\textbf{18.1\%} & Added L changes subset little \\
High-curatability & $P\geq2, R\geq3, L\geq1$ & 136 / 2,214 & \cellcolor{TableOrange}\textbf{6.1\%} & Joint evidence bottleneck \\
Preservation-review & $P\geq2, R\geq4, L\geq1$ & 50 / 2,214 & \cellcolor{TableRed}2.3\% & Multi-asset subset \\
\bottomrule
\end{tabularx}
\renewcommand{\arraystretch}{1.0}
\end{table*}

%% file: sections/05_results.tex
\section{Results}
\label{sec:results}

The results are organized around a visibility-to-curatability funnel. 
This results framework evaluates public traces as granular, distinct curation claims regarding artifact identity, scholarly linkage, upstream evidence, and release packages.
RQ1--RQ4 analyze these evidence fields separately, and RQ5 evaluates how often 
they co-occur in records coordinated enough for curation. 
Table~\ref{tab:key-results} summarizes the headline evidence components and 
readiness thresholds.

\begin{figure*}[!tbp]
    \centering
    \includegraphics[width=\textwidth]{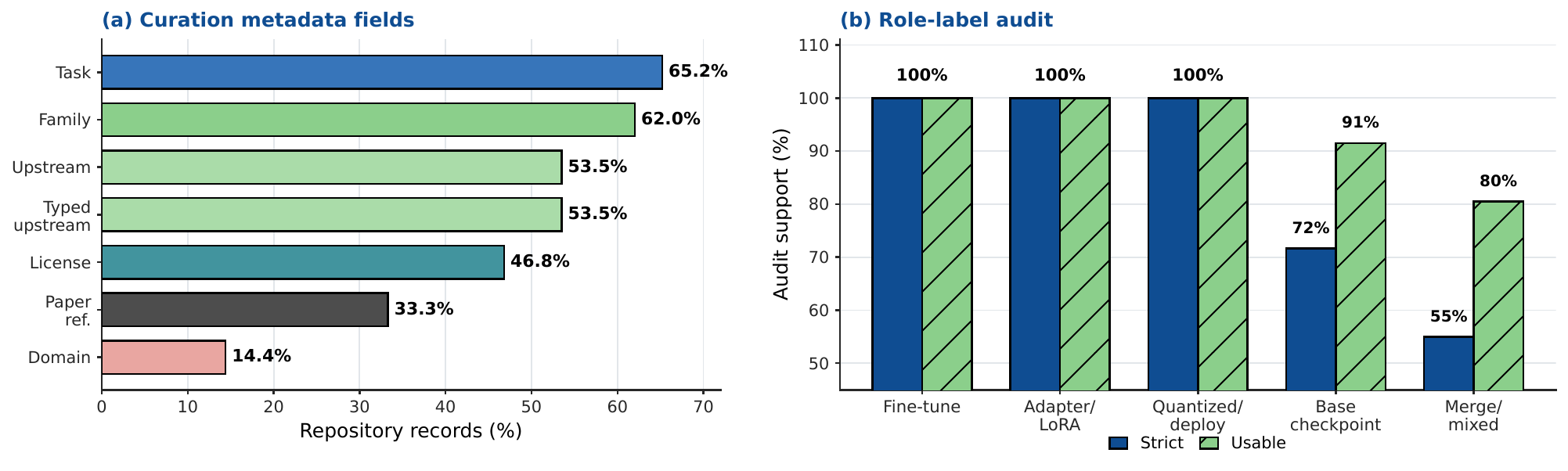}
    \caption{Repository-side metadata coverage and role audit. Panel (a) reports curation-relevant metadata-field coverage across the fixed repository snapshot. Panel (b) shows balanced-audit support for role labels, separating strict support from usable support. Fine-tune, adapter, and quantized/deployment labels are stable in the audit sample; base and merge/mixed labels have lower strict support and wider ambiguity gaps.}
    \Description{A two-panel figure. The first panel is a horizontal bar chart of metadata-field coverage in repository records. The second panel is a grouped vertical bar chart comparing strict and usable audit support for five artifact-role labels.}
    \label{fig:metadata}
\end{figure*}

\subsection{RQ1: Can Repository Records Identify Artifact Roles?}

Repository records provide broad coverage for artifact typing, but confidence is uneven across roles. In the fixed \hf snapshot, our pipeline assigns project-inferred artifact-role and package-form labels to all 191,375 repository records from public metadata and tags. According to the metadata coverage analyzed in Figure~\ref{fig:metadata}a, task descriptors, family labels, upstream fields, and license signals emerge as highly visible components, whereas scholarly references and domain descriptors remain sparse. Figure~\ref{fig:metadata}b shows that role typing is stable for several derived-artifact classes, while base and merge/mixed records require richer package-form information.

The balanced 300-repository expert audit finds 85.3\% overall strict role support. In the balanced sample, fine-tune, adapter/LoRA, and quantized/deployment records each reach 100.0\% strict support. Base records have 71.7\% strict support, and merge/mixed records have 55.0\% strict support. The inherent ambiguity of these boundary classes stems from overlapping signals where a single repository simultaneously exhibits base-family, merge, converted, or deployment characteristics, thereby rendering a generic model label insufficient to represent the underlying package structure.
\begin{figure*}[!tbp]
    \centering
    \includegraphics[width=\textwidth]{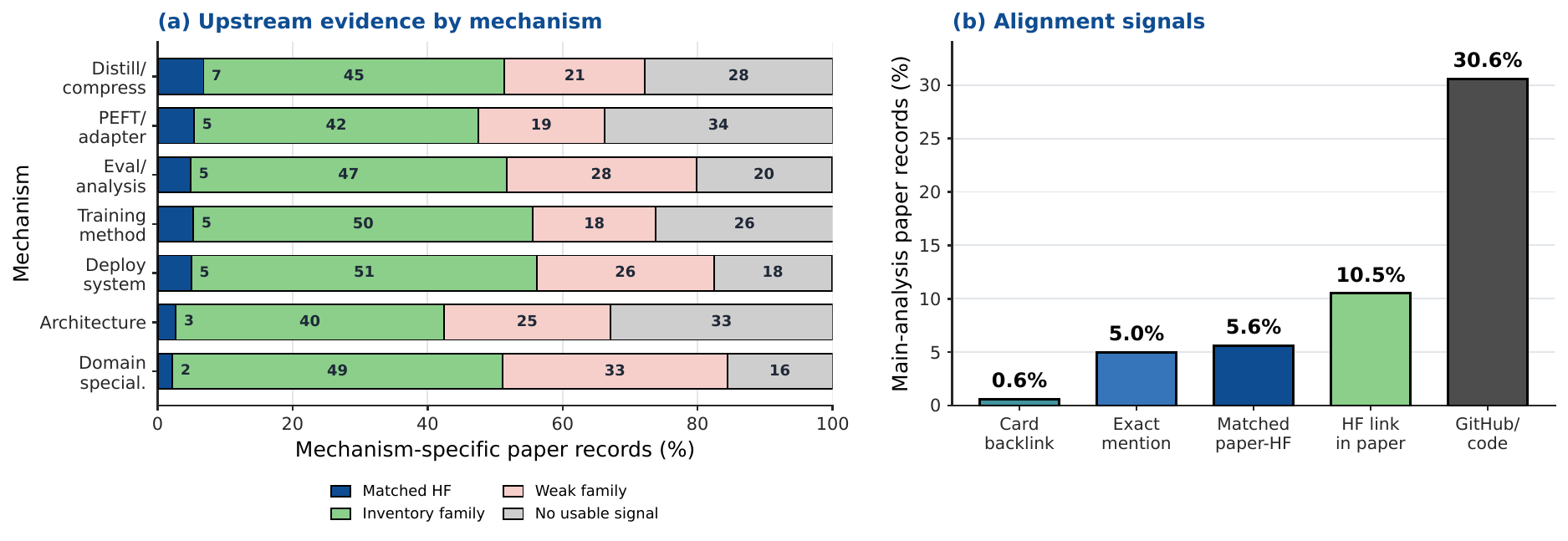}
    \caption{Upstream-evidence tiers and paper--repository alignment signals. Panel (a) reports the strongest public-record anchor available for upstream or family-level interpretation by mechanism. Panel (b) ranks direct alignment signals for bibliographic linkage. Alignment signals support provenance only when the linked paper or repository exposes an upstream relation.}
    \Description{A two-panel figure. The first panel is a set of horizontal stacked bars showing upstream and family evidence tiers by research mechanism. The second panel is a vertical bar chart separating paper text links, matched model mentions, model-card backlinks, and the combined matched-link measure.}
    \label{fig:provenance}
\end{figure*}
For curation, the result is not that every repository has a single unambiguous identity. It is that public metadata can support a first typing layer if catalogs preserve both the primary artifact role and the package form, rather than collapsing base checkpoints, adapters, quantized files, merged models, and deployment packages into one generic model label. The next question is whether those typed repository records can be connected to the papers that introduce, use, or evaluate them.

\subsection{RQ2: What Bibliographic and Linkage Evidence Is Visible?}

The linkage break occurs between visible links and typed scholarly relations. GitHub or code-hosting links appear in 30.6\% of analysis-set paper records and \hf links in 10.5\%, yet matched paper-to-\hf repository evidence covers only 5.6\%. The gap reflects relation typing and resolvability, since visible links may point to code, organizations, dependencies, examples, or auxiliary resources rather than the model-hub record introduced by the paper. The problem is therefore not the absence of links, but that links often lack a relation type.

Repository-side fields provide supporting context for linkage and discovery. Across the full snapshot, task descriptors appear in 65.2\% of repository records, model-family labels in 62.0\%, explicit upstream model fields in 53.5\%, typed upstream relation fields in 53.5\%, license tag signals in 46.8\%, scholarly references in 33.3\%, and domain descriptors in 14.4\%. These fields help organize a model collection, but they do not by themselves establish whether a repository is the artifact introduced by a paper, a dependency used by it, or a baseline cited for comparison.

Public traces therefore show that papers and repositories are frequently connected in some form, but these connections often remain fundamentally vague. A curatable record must resolve this ambiguity by establishing a typed scholarly relation that defines whether a repository is introduced by the paper, used as a dependency, evaluated as a baseline, cited as background, or provided as an official release package. Once this relation is typed, the next curatorial question is what the public record actually substantiates about the upstream origin of the artifact (RQ3) and the concrete assets provided in the release package (RQ4).

\begin{figure*}[!tbp]
    \centering
    \includegraphics[width=\textwidth]{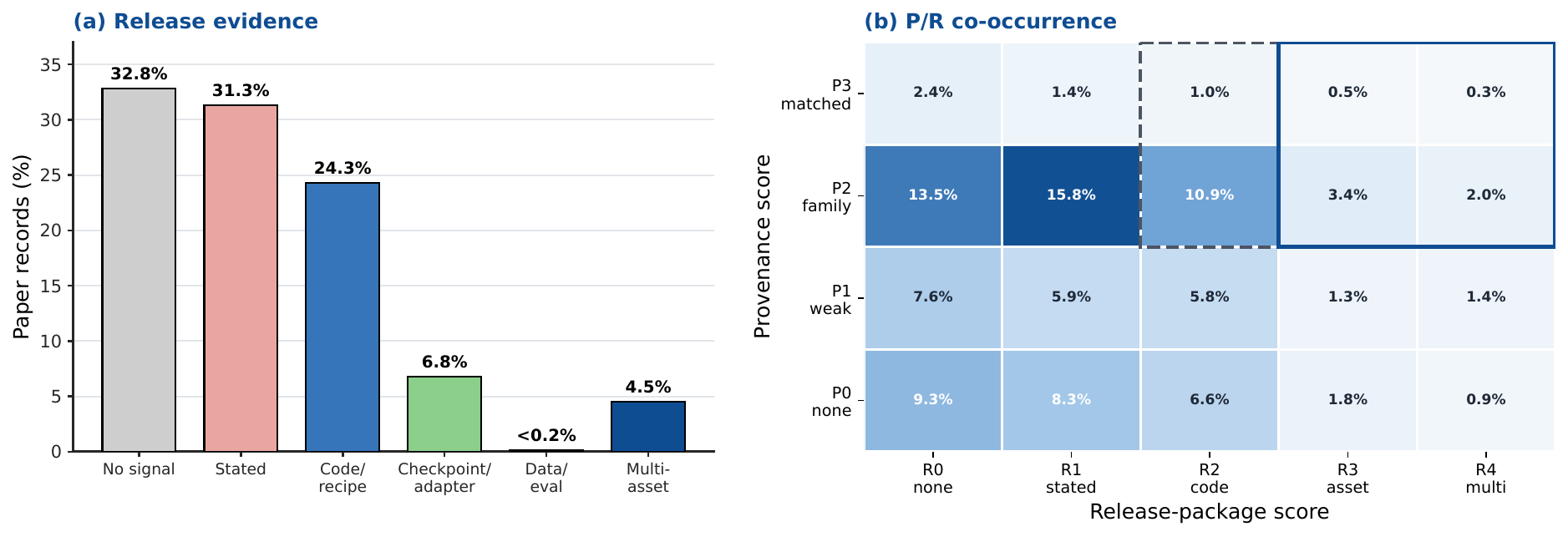}
    \caption{Release evidence and provenance--release co-occurrence. Panel (a) shows release-package evidence by category. Panel (b) cross-tabulates provenance score $P$ and release-package score $R$ for the 2,214 analysis-set paper records; the dashed box marks the operational P+R region and the solid box marks the asset-level region used by the high-curatability threshold.}
    \Description{A two-panel figure. The first panel is a vertical bar chart for release evidence tiers. The second panel is a heatmap crossing provenance scores with release-package scores and highlighting the operational and high-curatability threshold regions.}
    \label{fig:release}
\end{figure*}

\subsection{RQ3: What Upstream or Family-Level Evidence Can Public Records Support?}

Public records often support family-level collection organization, while exact upstream claims require typed relation evidence. Figure~\ref{fig:provenance} separates upstream-evidence tiers from paper--repository alignment signals. Panel (a) shows that this pattern holds across mechanisms: exact matched paper--\hf anchors are consistently rare, while inventory-backed or weak family evidence is much more common. Panel (b) decomposes direct alignment signals used for bibliographic linkage, which are not treated as provenance unless the linked record exposes an upstream relation.

Inventory-backed family signals cover 45.6\% of analysis-set paper records, weak or contextual family clues cover 21.9\%, and 26.9\% have no usable family or provenance signal. The expert upstream-evidence audit clarifies these signals: 28.0\% of 200 reviewed rows provide strict evidence for one clear primary upstream family, while 79.0\% provide usable family-level support for collection-level grouping. In curatorial terms, family evidence is often good enough for shelf placement, but not for a provenance claim.

The result supports a tiered upstream-evidence record. While family names are valuable for discovery and aggregate organization, a curatable record must preserve the specific relational role sustained by that family, whether as a base checkpoint, teacher, adapter target, merge input, quantization source, evaluator, baseline, or background context. Upstream evidence identifies what relation can be asserted, with release evidence determining what part of that artifact can actually be inspected or preserved.

\subsection{RQ4: What Release Package Evidence Is Visible?}

Release evidence is common at the statement level but rare at asset-package level. Figure~\ref{fig:release}a reports release-package distribution for 2,214 analysis-set paper records. Overall, 32.8\% have no explicit release signal, 31.3\% have a stated or incomplete release signal, 24.3\% provide code, scripts, or recipe evidence, 6.8\% provide checkpoint or adapter evidence, less than 0.2\% provide data or evaluation asset evidence, and 4.5\% provide multiple assets or a full recipe.

Collapsed into broader categories, 67.2\% contain some release or open-asset signal, 35.8\% contain concrete code or asset evidence, and 4.5\% contain strict multi-asset or full-recipe evidence. As a targeted boundary check, a 50-row expert URL/asset-level audit found that all reviewed URL/asset assignments matched the visible public-evidence category, thereby obviating the need for any aggregate category modifications.

Release should be recorded as a package, not as a binary availability flag. A paper can state that resources are released or expose GitHub scripts while still omitting weights, adapters, data, or evaluation assets, thereby helping inspection without providing a functional asset-level package. These tiers measure visible release evidence rather than role-specific release sufficiency. The final measurement evaluates the joint frequency of provenance and release evidence within a single paper record.

\subsection{RQ5: How Often Do the Required Evidence Fields Co-occur?}

The main bottleneck is combined curation readiness. Under the high-curatability threshold, a paper record must have usable provenance (P), asset-level or multi-asset release evidence (R), and at least one direct linkage signal (L). Only 136 of 2,214 analysis-set paper records satisfy this joint-evidence requirement, or 6.1\%.

Table~\ref{tab:key-results} reports the full threshold ladder. A broad discovery-ready rule covers 90.7\% of records. The operational P+R threshold, requiring usable provenance and code/recipe-or-stronger release evidence, covers 18.1\%. Adding the direct-link requirement changes this subset only marginally, from 401 to 400 records. The high-curatability threshold covers 6.1\%, and the preservation-review threshold requiring multi-asset or full-recipe evidence covers 2.3\%. Figure~\ref{fig:release}b shows where the drop occurs in the provenance--release co-occurrence map: the operational region with $P\geq2$ and $R\geq2$ contains 401 records, but the asset-level region with $P\geq2$ and $R\geq3$ contains only 136 records.

The key diagnostic is the operational P+R subset. Among 401 records with both usable provenance and concrete release evidence, 400 also contain at least one direct linkage signal. This is not a claim that release and linkage evidence are statistically independent, as a single paper-side code link can contribute both release evidence and direct alignment evidence. Instead, the operational+link check reveals that the relation gate is nearly saturated within this subset. Its small marginal effect is an empirical result rather than conceptual redundancy: $R$ records which release assets are visible, whereas $L$ records how a repository or asset relates to the focal paper. The substantial decline from 401 operational P+R records to 136 high-curatability records is driven by the release threshold, where 265 operational records expose code, scripts, or recipes without providing an asset-level package such as a checkpoint, adapter, data, evaluation asset, or multi-asset recipe. In this corpus, the main bottleneck is not an isolated missing link after provenance and concrete release evidence are present. It is the co-occurrence of usable provenance with sufficiently concrete release assets.

\input{tables/table_minimal_curatable_record}

Crucially, while paper records for open \llm artifacts are broadly visible, only a much smaller share expose enough coordinated evidence for provenance-aware curation. The funnel should therefore be read as an evidence-integration problem rather than an artifact-availability problem, in that the artifacts are often visible, yet the typed record fields needed to justify citation, attribution, reuse assessment, and preservation triage remain uncoordinated.

%% file: tables/table_minimal_curatable_record.tex
\begin{table*}[!tbp]
\centering
\caption{A minimal curatable record for compact and derived open \llm artifacts. The record acts as a coordination contract across existing systems rather than a requirement for a new centralized registry.}
\label{tab:minimal-curatable-record}
\small
\setlength{\tabcolsep}{4pt}
\renewcommand{\arraystretch}{1.12}
\begin{tabularx}{0.98\textwidth}{@{}p{0.15\textwidth}p{0.20\textwidth}p{0.22\textwidth}p{0.20\textwidth}Y@{}}
\toprule
Field group & Core field & Primary maintainer/source & Failure prevented & Curation use \\
\midrule
Artifact identity & \texttt{artifact\_role} & Model hub / repository owner & Generic model label & Catalog type \\
\rowcolor{TableGray}
Package form & \texttt{package\_form} & Model hub / repository metadata & Hidden distribution format & Package interpretation \\
Scholarly relation & \texttt{paper\_repository\_relation} & Paper authors + scholarly index & Untyped links & Bibliographic control \\
\rowcolor{TableGray}
Upstream relation & \texttt{upstream\_relation\_type} & Model hub + paper text & Family-as-provenance & Provenance relation \\
Evidence tier & \texttt{provenance\_evidence\_tier} & Digital library / index & Overstated confidence & Confidence and scope \\
\rowcolor{TableGray}
Release assets & \texttt{release\_asset\_types} & Model hub + code repo + paper & Release-as-binary & Asset package \\
License/access signal & \texttt{license\_access\_signal} & Model hub / repository metadata & Missing governance context & Access context \\
\bottomrule
\end{tabularx}
\renewcommand{\arraystretch}{1.0}
\end{table*}

%% file: sections/06_discussion.tex
\section{Discussion}
\label{sec:discussion}

The results show that compact and derived open \llm{} artifacts have entered the public scholarly record, but not yet as well-formed curation records. A curator looking at a paper and its public traces needs to answer four practical questions: What is the artifact? Which paper relation does the repository have? What upstream relation is evidenced? What assets can be inspected or preserved? The visibility-to-curatability funnel shows that these answers often exist separately, but rarely appear together in one typed record. This section translates that gap into a minimal record design.
\subsection{From Visibility to a Minimal Curatable Record}

Table~\ref{tab:minimal-curatable-record} summarizes the resulting minimal 
curatable record. The design is intentionally minimal. Rather than introducing a comprehensive new metadata schema, it selects only the fields necessary to block the curation ambiguities observed in the funnel and audits. Artifact role 
and package form resolve identity collisions by specifying what the artifact is 
and how it is distributed. Scholarly relation links the artifact to specific 
literature. Upstream relation and provenance-evidence tier separate casual family 
mentions from stronger derivation claims. Release assets define the inspectable or preservable content, while licensing and access signals add governance context. The failure-prevention column makes the design logic explicit by aligning each field with a specific ambiguity observed in the funnel or audits.

The record operates as a coordination layer across existing infrastructure rather 
than as a centralized registry. It does not require a single authority to certify 
every open-model artifact. Instead, it leverages the same public traces that made 
our empirical measurement possible, while keeping their sources and evidentiary 
strength explicit. While these traces are currently scattered across model hubs, paper records, code repositories, and scholarly indexes, the minimal record functions as a coordination schema that keeps them joinable without collapsing them into a single, undifferentiated availability label. The goal is not to impose a heavier publication burden, but to make 
existing public traces usable for citation, attribution, reuse assessment, and 
preservation triage.

\paragraph{Worked construction.}
Table~\ref{tab:worked-case} applies the seven fields to MobileLLM-R1-360M, a sub-billion-parameter reasoning model represented in the audited corpus~\cite{zhao2026mobilellmr1}. Three evidence bundles connect its identity and paper relation, exact upstream evidence, and multi-asset release into an auditable status.
\input{tables/table_worked_case}

The audit boundary cases reinforce the same design. Family names may denote targets, parents, teachers, baselines, or background, while code links may expose scripts without weights. A curatable record must preserve these relation types and evidence strengths rather than compressing them into a generic ``open model'' label.

\subsection{Infrastructure Responsibilities and Design Implications}

The funnel also indicates where missing evidence is most likely to be maintained. 
Because the main drop occurs when curation requires asset-level or multi-asset 
release evidence, artifact-role, package-form, release-asset, and license/access 
fields should be maintained closest to model repositories. Upstream-relation 
fields require both model-hub declarations and paper context, because a family 
signal may refer to a base checkpoint, teacher model, adapter target, merge input, 
quantization source, evaluator, baseline, or background mention. Because typed 
paper--repository relations are sparse overall and raw links often remain untyped, 
paper authors and scholarly indexes are closest to paper--repository relation 
types. Because family evidence is often usable for grouping but weak for strict 
lineage claims, digital libraries and preservation systems should preserve 
provenance tiers and curation-readiness status across sources. The coordination layer focuses on maintaining the joinability of public traces while preserving their source context and evidentiary strength, bypassing the need for a centralized registry.

This division of labor leads to evidence-to-action interventions.
\begin{enumerate}[leftmargin=*,itemsep=2pt,topsep=2pt]
    \item \textbf{Role ambiguity $\rightarrow$ separate artifact role from package form.} Model hubs should distinguish base checkpoints, fine-tunes, adapters, merges, quantized files, and deployment packages without forcing them into one generic model label.
    \item \textbf{Untyped linkage $\rightarrow$ type paper--repository relations.} Paper authors and scholarly indexes should distinguish repositories introduced by a paper, used as dependencies, evaluated as baselines, cited as background, or provided as official release packages.
    \item \textbf{Family-as-provenance risk $\rightarrow$ type upstream relations.} Model hubs and paper records should distinguish base, teacher, adapter-target, merge-input, quantization-source, evaluator, baseline, and background-mention relations.
    \item \textbf{Uneven evidence strength $\rightarrow$ represent provenance tiers.} Digital libraries should preserve whether a claim rests on an explicit upstream field or statement, a matched repository with a visible upstream relation, an inventory-backed family signal, or a contextual mention.
    \item \textbf{Binary release labels $\rightarrow$ expose release packages as structured fields.} Release records should separate code, scripts, weights or checkpoints, adapters, datasets, evaluation assets, recipes, and environment details.
\end{enumerate}

These interventions align with FAIR software and research-object packaging work, 
which treats reuse as a property of structured packages and relations rather than 
of a single access flag \cite{barker2022fair4rs,soilandreyes2022rocrate}. The 
Model Openness Framework makes a related point at the level of openness 
components \cite{white2024mof}. The contribution here is an empirical, record-level analysis demonstrating which curation evidence fields are visible, which remain weak, and which fail to co-occur within the same public record. Digital-library 
infrastructure should therefore make the evidence needed for citation, provenance 
tracking, reuse assessment, and preservation selection explicit, typed, and 
actionable.

%% file: tables/table_worked_case.tex
\begin{table*}[!tbp]
\centering
\caption{Compact worked example for MobileLLM-R1-360M. Three evidence bundles populate the seven-field record and integrated status.}
\label{tab:worked-case}
\small
\setlength{\tabcolsep}{5pt}
\renewcommand{\arraystretch}{1.05}
\begin{tabularx}{0.98\textwidth}{@{}>{\raggedright\arraybackslash}p{0.16\textwidth}>{\raggedright\arraybackslash}p{0.35\textwidth}Y@{}}
\toprule
Evidence bundle & Example public trace & Typed record \\
\midrule
Identity + relation & Official model card identifies a 360M fine-tuned model and cites the focal paper & \texttt{artifact\_role}: fine-tuned model; \texttt{package\_form}: checkpoint; \texttt{paper\_repository\_relation}: official release ($L=2$) \\
\rowcolor{TableGray}
Provenance & Model tree identifies MobileLLM-R1-360M-base as the upstream checkpoint & \texttt{upstream\_relation\_type}: fine-tuned from; \texttt{provenance\_evidence\_tier}: explicit/canonical ($P=3$) \\
Release + status & Checkpoints, code, recipes, data sources, and a research-use license are visible & \texttt{release\_asset\_types}: multi-asset package; \texttt{license\_access\_signal}: public, research use; $P=3$, $R=4$, $L=2$: preservation-review candidate \\
\bottomrule
\end{tabularx}
\renewcommand{\arraystretch}{1.0}
\end{table*}

%% file: sections/07_limitations_conclusion.tex
\section{Scope and Limitations}
\label{sec:limitations}

This study measures public-record sufficiency, not model quality, adoption, complete genealogy, executable reproducibility, or preservation success. The repository collection is a scoped May 2026 \hf snapshot, and paper-level results use the 2,214 analysis-set paper records unless otherwise stated. Artifact roles, package forms, provenance tiers, and release-package levels are derived from visible records and calibrated through audits, so private training logs, hidden checkpoints, unreported assets, and developer knowledge remain outside scope. Release tiers measure visible asset evidence rather than role-specific package sufficiency. Generative-AI tools assisted with candidate evidence extraction from unstructured paper text, language editing, and code/script drafting, while all final analytical fields, claims, released tables, and manuscript text were verified by the authors.

\section{Conclusion}
\label{sec:conclusion}
This study examines the curatability of compact and derived open \llm{} artifacts from a digital library perspective. We introduced a record-level framework for operationalizing curatability and conducted a collection-scale analysis of 191,375 Hugging Face repositories and 2,214 scholarly papers. Our findings reveal a pronounced visibility-to-curatability funnel: although 90.7\% of paper records contain at least one useful curation signal, only 6.1\% provide sufficiently coordinated evidence to support provenance-aware curation.
These results suggest that the primary challenge for preserving open \llm{} artifacts is not resource availability, but evidence coordination across distributed scholarly records. We therefore propose a minimal curatable record that preserves artifact identity, scholarly linkage, provenance, and release assets while maintaining their evidentiary strength and source context. As AI research continues to produce increasingly interconnected computational artifacts, digital libraries will need to move beyond cataloging individual repositories or publications toward coordinating evidence across the broader scholarly ecosystem.

\section{Disclosure of AI Use}
The authors used generative AI tools to assist with language editing, improve the clarity of the manuscript, and support code drafting and software development. All AI-generated code was reviewed, tested, and validated by the authors before use in this study. The authors developed, verified, and take full responsibility for all scientific content, analyses, interpretations, and conclusions presented in this manuscript.